\documentclass[runningheads]{llncs}

\usepackage{graphicx}
\usepackage{amsfonts,latexsym}
\usepackage{amsmath}
\usepackage{amssymb}
\usepackage{enumitem}
\usepackage{url}

\usepackage{listings}
\usepackage{xcolor}
\definecolor{codegreen}{rgb}{0,0.6,0}
\definecolor{codegray}{rgb}{0.5,0.5,0.5}
\definecolor{codepurple}{rgb}{0.58,0,0.82}
\definecolor{backcolour}{rgb}{0.95,0.95,0.92}
\definecolor{magnolia}{rgb}{0.97, 0.96, 1.0}
\definecolor{azure}{rgb}{0.94, 1.0, 1.0}
\definecolor{persianred}{rgb}{0.8, 0.2, 0.2}
\definecolor{persianblue}{rgb}{0.11, 0.22, 0.73}
\definecolor{newbackground}{rgb}{0.98, 0.98, 0.98}

\lstdefinestyle{mystyle}{
	backgroundcolor=\color{newbackground},   
	commentstyle=\color{codegreen},
	keywordstyle=\color{magenta},
	numberstyle=\tiny\color{codegray},
	stringstyle=\color{persianblue},
	basicstyle=\ttfamily\footnotesize,
	breakatwhitespace=false,         
	breaklines=true,                 
	captionpos=b,                    
	keepspaces=true,                 
	numbers=left,                    
	numbersep=5pt,                  
	showspaces=false,                
	showstringspaces=false,
	showtabs=false,                  
	tabsize=2,
	frame=lines
}

\lstdefinelanguage{Maple}
{morekeywords={proc, do, end, proc, for, from, to, then, package, return, module, object, return, local, export, error, if},
	morecomment=[l]{\#}
}
\usepackage{algorithm}
\usepackage{algorithmicx}
\usepackage{algpseudocode}
\usepackage{verbatim}

\usepackage[belowskip=-15pt,aboveskip=0pt]{caption}

\newcommand{\HomoPart}[2]{\mbox{${#1}_{(#2)}$}}
\newcommand{\Truncate}[2]{\mbox{${#1}^{(#2)}$}}

\newcommand{\totaldegree}[1]{| #1 |}
\newcommand{\order}[1]{\mbox{{\rm ord}$(#1)$}}
\newcommand{\Mk}[1]{\mbox{$\mathcal{M}^{#1}$}}
\newcommand{\MPS}{\tt Multivariate\-Power\-Series}
\newcommand{\PSO}{\tt Power\-Series\-Object}
\newcommand{\UPoPSO}{\tt Univariate\-Polynomial\-Over\-Power\-Series\-Object}
\newcommand{\Sage}{{\sc  SageMath}}

\def\KXN {\ensuremath{\mathbb{K}[\![ X_1, \ldots, X_n ]\!]}}
\def\M {\ensuremath{\mathcal{M}}}

\newcommand{\EAO}{\mbox{\sc EvaluateAtOrigin}}
\newcommand{\HF}{\mbox{\sc HenselFactorize}}
\newcommand{\WP}{\mbox{\sc WeierstrassPreparation}}
\newcommand{\TS}{\mbox{\sc TaylorShift}}

\def\A {\ensuremath{\mathbb{A}}}

\def\C {\ensuremath{\mathbb{C}}}

\def\K {\ensuremath{\mathbb{K}}}

\def\N {\ensuremath{\mathbb{N}}}
\def\Q {\ensuremath{\mathbb{Q}}}
\def\R {\ensuremath{\mathbb{R}}}
\newcommand{\T}{\mathfrak{T}}

\newcommand{\F}{\mathcal{F}}

\newcommand{\Maple}{{\sc  Maple}}

\newcommand{\BPAS}{{\sc  BPAS}}

\newcommand{\maple}{{\sc  Maple}}

\newcommand{\RegularChains}{{\tt  Regu\-lar\-Chains}}

\newcommand{\todo}[2]{{{\textcolor{red}{ #1}}}\footnote{ {\textcolor{blue}{ #2}} }}

\newcommand{\fixed}[3]{#1}
\newcommand{\reworked}[2]{#2}
\newcommand{\removed}[3]{}

\begin{document}
\title{Multivariate Power Series in Maple}
\author{Mohammadali Asadi\inst{1, 2} \and Alexander Brandt \inst{1} \and 
Mahsa Kazemi \inst{1} \and Marc Moreno Maza \inst{1} \and Erik Postma\inst{2}}
\authorrunning{M. Asadi et al.}
\institute{The University of Western Ontario, 1151 Richmond St, London, Ontario, Canada \and
Maplesoft, 615 Kumpf Dr, Waterloo, Ontario, Canada}
\maketitle
\begin{abstract}
  \reworked{We discuss our implementation of power series 
which is available as {\MPS} in {\Maple} 2021.
This library provides a variety of methods
to study formal power series and univariate 
polynomials over power series (UPoPS) by
utilizing object-oriented mechanisms and 
lazy evaluation techniques. The library is written using
{\Maple} classes and provides a simple and easy-to-use user interface. 
We report on the performance of our implementation of 
lazy arithmetic for multivariate power series,
as well as Weierstrass Preparation Theorem and 
factorization via Hensel's lemma for UPoPS objects.}{
    We present {\MPS}, a {\Maple} library introduced in {\Maple} 2021, providing a variety of
    methods to study formal multivariate power series and univariate polynomials
    over such series.
    This library offers a simple and easy-to-use user interface.
    Its implementation relies on lazy evaluation techniques and 
    takes advantage of
    {\Maple}'s features for object-oriented programming.
    The exposed methods include Weierstrass Preparation Theorem and 
factorization via Hensel's lemma.
    The computational performance is demonstrated
    by means of an experimental comparison with
    software counterparts.
}

\keywords{multivariate power series \and Weierstrass Preparation Theorem \and Hensel's lemma \and factorization \and lazy evaluation}
\end{abstract}

	\section{Introduction}
In elementary courses on univariate calculus, 
power series are often introduced as limits of sequences 
of the form ``the first \reworked{$n \in \N$}{$ n $} terms of a given sequence''. 
This leads students to the study of analytic functions and 
the use of power series in computing function limits. 
While the extension of those notions to the multivariate case 
is a standard topic in advanced calculus courses, 
the availability of multivariate power series and multivariate 
analytic functions in computer algebra systems is somehow limited.
In {\sc Maple} \cite{maple}, {\sc SageMath} \cite{sageMath}, and {\sc Mathematica} \cite{mathematica}, 
\fixed{power series are restricted to being either only univariate or truncated,}
{which are restricted to univariate? which are restricted to truncated?}
{Maple and SAGE have both, I do not know about Mathematica.}
that is, reduced modulo a fixed power of the ideal 
$\langle X_1, \ldots, X_{n} \rangle$ 
generated by the variables of those power series. 
A truncated implementation, while simple, may be insufficient for, 
or computationally more expensive in, some particular circumstances. 
For instance, modern algorithms for polynomial system solving require
the intensive use of modular methods based on Hensel lifting. 
In those lifting procedures,
degrees of truncation may not be known a priori, 
thus leading to truncated power series being ineffective. 
Considering that a power series has potentially an infinite number of 
terms naturally suggests to represent it as a procedure which,
given a particular (total) degree, produces the terms of that degree. 
This leads to a so-called lazy evaluation 
scheme,
where the terms of any power series
are produced only as needed, 
via such a \textit{generator} function.
The usefulness of lazy evaluation in computer algebra has been studied for
a few decades. In particular, see the work of Karczmarczuk \cite{karczmarczuk1997generating},
discussing different mathematical objects with an infinite length;
Burge and Watt \cite{burge1987infinite}, and van der Hoeven \cite{van2002relax}, discussing 
lazy univariate power series; and Monagan and Vrbik \cite{DBLP:conf/casc/MonaganV09},
discussing lazy arithmetic for polynomials. 
In this paper, 
we present {\MPS}, which is  among the new features released in {\Maple} 2021 and publicly available in \cite{MPSgithub}.
This library, written in the {\Maple} language, provides 
the ability to create and manipulate 
multivariate power series \reworked{over numeric and algebraic coefficient rings,}
{with rational or algebraic number coefficients,} 
as well as univariate polynomials whose coefficients are
multivariate power series.
Through lazy evaluation techniques and a careful implementation,
our library achieves very high performance. 
These power series and univariate polynomials over power series (UPoPS)
are employed in optimized implementations of 
Weierstrass Preparation Theorem and factorization 
of UPoPS via Hensel's lemma.
Our implementation follows the lazy evaluation scheme of multivariate power series 
in the {\BPAS} library \cite{bpasweb}.
The multivariate power series of {\BPAS}, written in the C language, is discussed in \cite{brandt2020power}
and extends upon the work of the {\tt PowerSeries} subpackage
of the {\tt RegularChains} {\Maple} library \cite{alvandi2016computing,Marc18LecNote}. 
The {\tt PowerSeries} package is the only preexisting implementation of multivariate power series integrated in {\Maple}.
In \cite{brandt2020power}, it is shown that the {\BPAS} implementation provides exceptional
performance, surpassing that of the {\tt PowerSeries} package, the basic {\Maple} function {\tt mtaylor}, 
and the multivariate power series available in SageMath \cite{sageMath} by multiple orders of magnitude. 
A key design element of our library, in addition to lazy evaluation techniques, 
is the use of {\Maple} {\em objects} and object-oriented programming.
An object in {\Maple} is a special kind of 
module
which encapsulates together data and procedures 
manipulating that data, just like objects in any 
other object-oriented language; see \cite[Chapters 8, 9]{mapleProgGuide}.
To the best of our knowledge, 
few {\Maple} libraries make use of those objects, which, 
as our report suggests, are worth considering for improving performance. 
In particular, objects allow for the overloading of existing builtin {\Maple} functions
in order to integrate these new custom objects with existing {\Maple} library code.
Our results show that {\MPS} is comparable in performance to the implementation of 
{\BPAS}, is thus similarly several orders of magnitude faster than other
existing implementations.
These experimental results are discussed
in Section~\ref{sec:exper}.
The remainder of this paper is organized as follows.  We begin in
Section~\ref{sec:bg} with reviewing definitions of formal power
series, and univariate polynomial over power series, followed by a
brief discussion about the basic arithmetic, Weierstrass preparation
theorem and factorization via Hensel's lemma. 
Section~\ref{sec:UI} presents an overview of the {\MPS} package, 
while Section~\ref{sec:princ} explores its underlying design principles.
Implementation details are discussed in Section~\ref{sec:impl},
followed by our experimentation in Section~\ref{sec:exper}.  Finally,
we conclude and present future works in Section~\ref{sec:conc}.
\section{Background}
\label{sec:bg}
In this section we review the basic properties of formal power series and 
univariate polynomials over those series, following G. Fischer in~\cite{fischerCurves}.
While various proofs of Theorems~\ref{WeierstrassTheorem}
of ~\ref{HenselLemma} can be found in the literature,
the proofs given in~\cite{brandt2020power} are constructive
and support our implementation.
Throughout this paper, $\N$ denotes the semi-ring of non-negative integers
and {\K} an algebraic number field.
\subsection{Power Series}
Given a positive integer $n$,
we denote by $\KXN$ 
the set of multivariate formal power series
with coefficients in {\K} and variables $X_1, \ldots, X_n$.
Let $f = \sum_{e \in {\N}^{n}} a_e X^e \in \KXN$ \fixed{and $d \in {\N}$}{degree $d$ can be $0$, however, in the CASC paper $d \in \N$,
  why?. AB: I think this is a result of the frequent ambiguity of
  $N=\{1,2,...\}$ or $N=\{0,1,...\}$. From my CS background I always
  think of $N=\{0,1,...\}$ and $N^{+}=\{1,2,\ldots\}$. I think that in
  more-modern and less-pure-math contexts $N=\{0,1,...\}$ is the more
  common case? But, then I don't know how defining $N=\{0,1,...\}$
  would change the definition of Cauchy sequences. Marc can weigh
  in.}{Marc: I agree with Alex} where  $X^e = X_1^{e_1} \cdots X_n^{e_n}$ and 
$e = (e_1, \ldots, e_n) \in {\N}^n$.
The {\it homogeneous part} and {\it polynomial part} of $f$ in degree $d$ are
respectively defined by 
$\HomoPart{f}{d} := \sum_{\totaldegree{e} = d} a_e X^e$ and 
$\Truncate{f}{d} := \sum_{k \leq d} \HomoPart{f}{k}$, where 
$\totaldegree{e} =  e_1 + \cdots + e_n$.
The sum (resp. difference) of two formal power series 
$f, g \in \KXN$ is defined by the sum (and resp. difference) 
of their homogeneous parts of the same degree; thus we have:
$f \pm g = \sum_{d \in {\N}} \HomoPart{f}{d} \pm \HomoPart{g}{d}$.
The product $h = f \cdot g$ can be defined as $h = \sum_{d \in {\N}} \HomoPart{h}{d}$ with $\HomoPart{h}{d} = \sum_{k + l = d} \HomoPart{f}{k} \ \HomoPart{g}{l}$. 
With the above addition and multiplication,
the set $\KXN$ is a local ring
with $\M :=  \langle X_1, \ldots, X_{n} \rangle$ as maximal ideal;
$\KXN$ is also a unique factorization domain (UFD).
The order of the power series $f$, denoted by \order{f}, is defined as 
$min\{ d \in {\N} \ \vert \ \HomoPart{f}{d} \neq 0 \}$ if $f \neq 0$, and as $\infty$ otherwise. 
We observe that $\Mk{k} = \{ f \in \KXN \ \vert \ \order{f} \geq k \}$
holds for every $k \geq 1$.
If $f$ is a unit, that is, if  $f \not\in \M$
(or equivalently,  if $\order{f} = 0$)
then the sequence $(h_m)_{m \in \N}$, where $h_m = c^{-1} (1 + g + \cdots + g^m)$,
$c = \HomoPart{f}{0}$, 
and $g = 1 - c^{-1} f$, converges to the {\em inverse} of $f$.
This convergence is the sense of Krull topology, see~\cite{fischerCurves}
for details.
\subsection{Univariate Polynomials over Power Series}

We denote by $\A$ and $\M$
the power series ring 
$\mathbb{K}[\![ X_1, \ldots, X_{n} ]\!]$ and
its maximal ideal.
We allow $n = 0$, in which case we have 
$\M = \langle 0 \rangle$.
Let $f \in {\A}[\![X_{n+1}]\!]$, written as $f  = \sum_{i=0}^{\infty} \,  a_i X_{n+1}^i$
with $a_i \in {\A}$ for all $i \in \N$.
Then, Weierstrass Preparation Theorem (WPT) states the following.
\begin{theorem} \label{WeierstrassTheorem}
Assume $f \not\equiv 0 \mod{ {\cal M}}[\![X_{n+1}]\!]$.
Let $d \geq 0$ be the smallest integer such that $a_d \not\in {\cal M}$.  
Then, there exists
a unique pair $({\alpha}, p)$ satisfying 
the following:
\begin{enumerate}[label=\roman*.]
\item ${\alpha}$ is an invertible power series of ${\A}[\![X_{n+1}]\!]$,
\item $p \in {\A}[X_{n+1}]$ is a monic polynomial of degree $d$,
\item writing $p = X_{n+1}^d + b_{d-1} X_{n+1}^{d-1} + \cdots + b_1 X_{n+1} + b_0$,
               we have $b_{d-1}, \ldots, b_0 \in {\cal M}$,
\item $f = {\alpha} p$ holds.
\end{enumerate}
Moreover, if $f$ is a polynomial of ${\A}[X_{n+1}]$ of degree $d+m$,
for some $m$, then ${\alpha}$ is a polynomial of ${\A}[X_{n+1}]$ of degree $m$.
\end{theorem}

Since $\A$ is a UFD, then Gauss' lemma implies that 
the polynomial ring $\A [X_{n+1}]$ is also a UFD.
Hensel's lemma shows how factorizing a
polynomial in $\A [X_{n+1}]$ can
be reduced to factorizing a polynomial
in $\mathbb{K}[X_{n+1}]$.
\begin{theorem}[Hensel's Lemma]\label{HenselLemma}
Assume that $f$ is a polynomial of degree $k$ in ${\A}[X_{n+1}]$.
We define $\overline{f} = f(0, \ldots, 0, X_{n+1}) \in {\K}[X_{n+1}]$.
We assume that $f$ is monic in $X_{n+1}$, that is, $a_k = 1$.
We further assume that ${\K}$ is algebraically closed.
Thus, there exists positive integers $k_1, \ldots, k_r$ and pairwise distinct elements $c_{1}, \ldots, c_{r} \in {\K}$ 
such that we have $\overline{f} = (X_{n+1} - c_1)^{k_1} (X_{n+1} - c_2)^{k_2}  \cdots (X_{n+1} - c_r)^{k_r}.$
Then, there exists $f_1, \ldots, f_r \in {\A} [X_{n+1}]$,
all monic in $X_{n+1}$, such that we have:
\begin{enumerate}[label=\roman*.]
\item $f = f_1 \cdots f_r$,
\item the degree of $f_j$ is $k_j$, for all $j = 1, \ldots, r$,
\item $\overline{f_j} = (X_{n+1} - c_j)^{k_j}$, for all $j = 1, \ldots, r$.
\end{enumerate}
\end{theorem}

    \section{An Overview of the User-Interface}
\label{sec:UI}

\begin{figure}[H]
 \centering
 \includegraphics[width=0.9\textwidth]{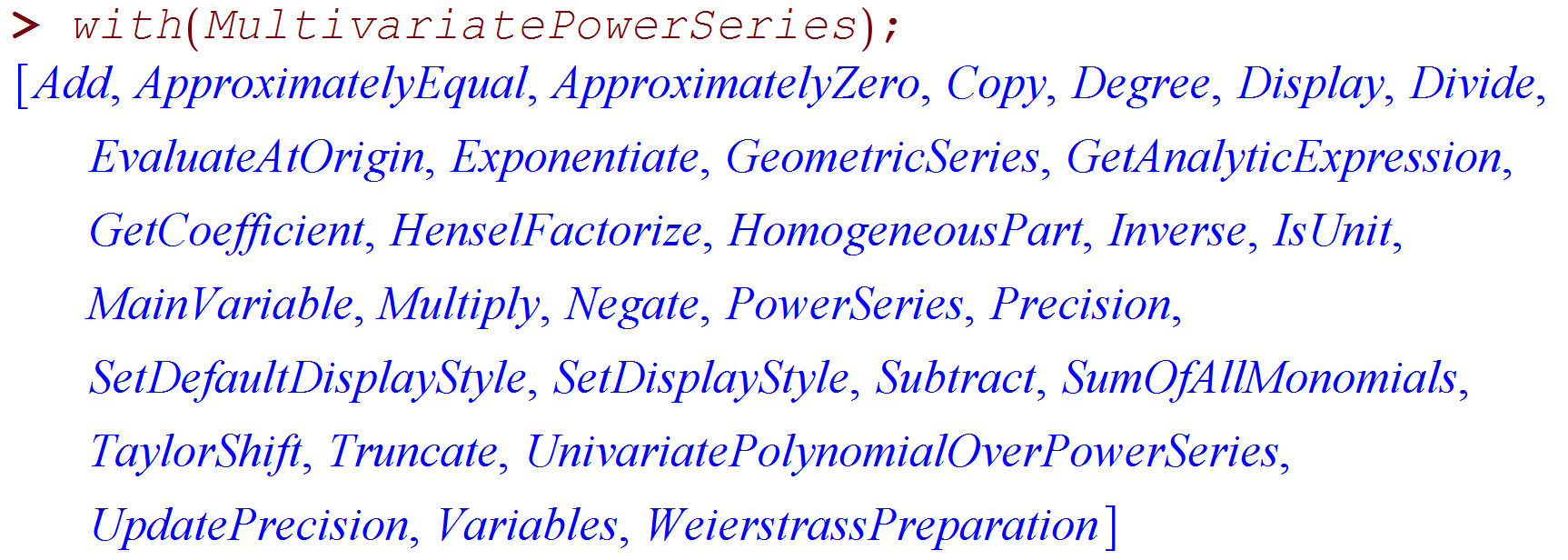}
 \caption{List of the commands of {\MPS}.}
   \label{fig:PowerSeries-0}
\end{figure}

From the point of view of the end-user, the {\MPS} package is a
collection of commands for manipulating multivariate power series and
univariate polynomials over multivariate power series.
The field of
coefficients of all power series created by the command {\tt
  PowerSeries} consists of all complex numbers that are constructible
in {\Maple}, thus including rational numbers and algebraic numbers.
The main algebraic functionalities of this package deal with
arithmetic operations (addition, multiplication, inversion,
evaluation), for both multivariate power series and univariate
polynomials over multivariate power series (UPoPS), as well as
factorization of such polynomials.  The list of the exposed commands
is given in Figure \ref{fig:PowerSeries-0}.

\begin{figure}[H]
 \centering
 \includegraphics[width=\textwidth]{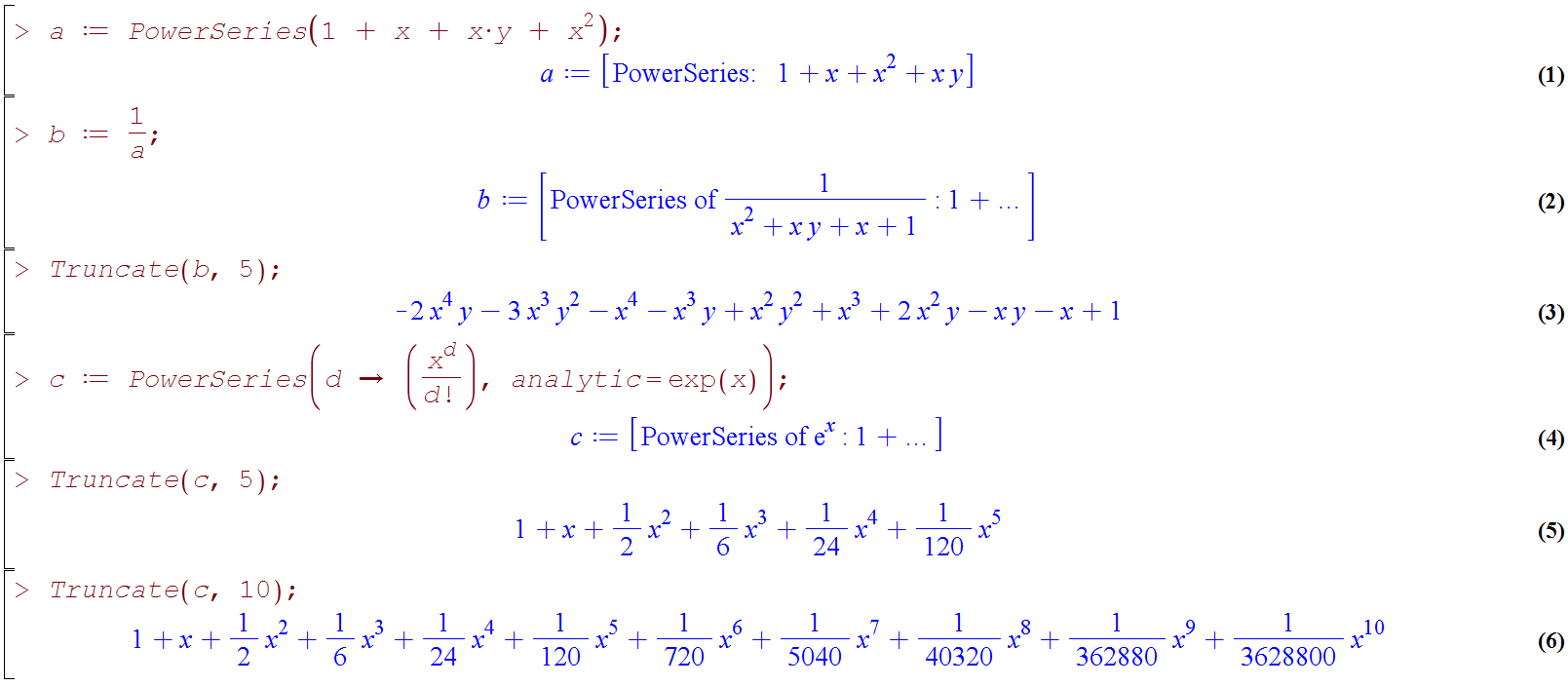}
 \caption{Creating power series from a polynomial or an anonymous function.}
   \label{fig:PowerSeries-1}
\end{figure}

\begin{figure}[H]
 \centering
 \includegraphics[width=\textwidth]{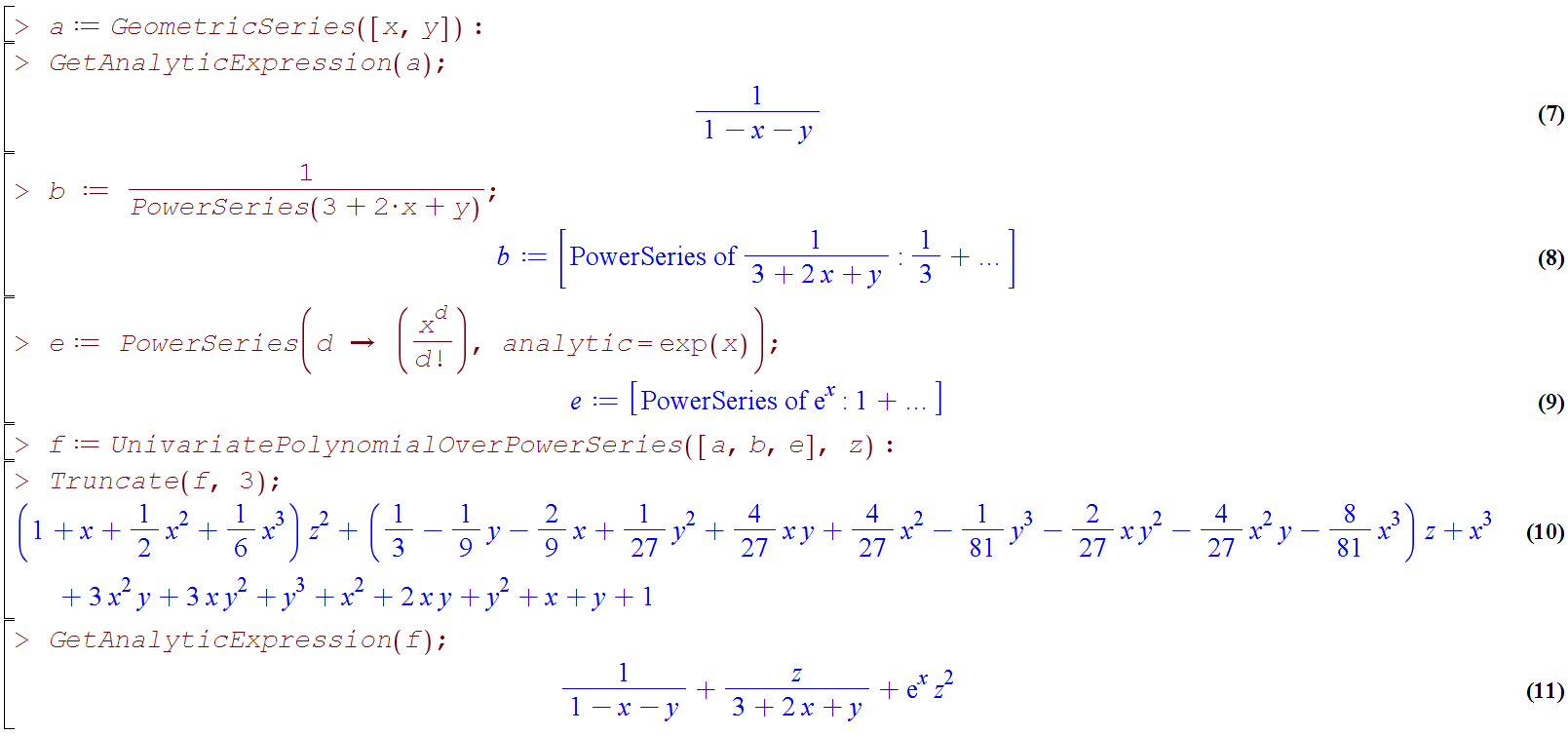}
 \caption{Creating a univariate polynomial over power series}
   \label{fig:PowerSeries-3}
\end{figure}

The commands {\tt PowerSeries} and {\tt UnivariatePolynomialOverPowerSeries}
create power series and univariate polynomials over multivariate power
series, respectively, from objects like polynomials, sequences, and functions which
produce homogeneous parts of a power series, 
as illustrated in Figures \ref{fig:PowerSeries-1}
and \ref{fig:PowerSeries-3}.
The commands {\tt GeometricSeries} and {\tt SumOfAll\-Monomials}
respectively create the geometric series and sum of all monomials for an input list of variables.

\begin{figure}[H]
	\centering
	\includegraphics[width=\textwidth]{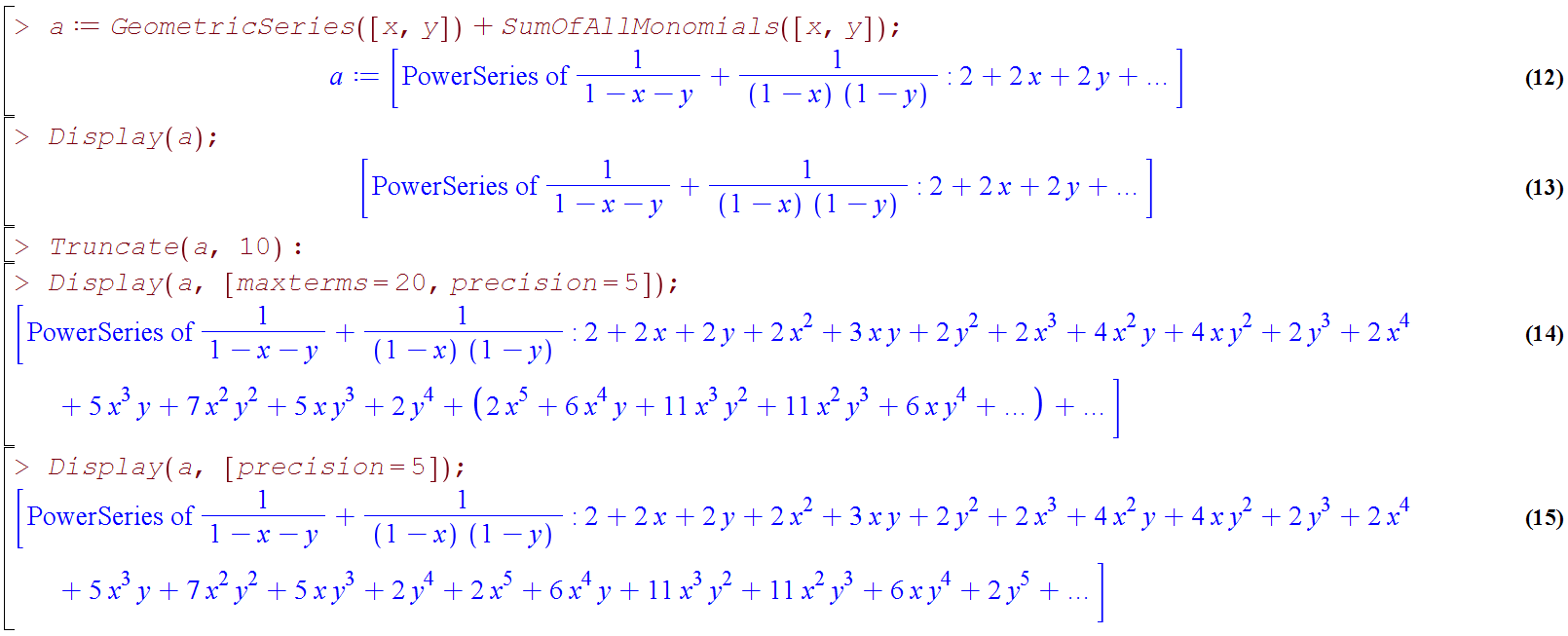}
	\caption{Controlling the output format of a multivariate power series.}
	\label{fig:PowerSeries-2}
\end{figure}

Whenever possible, the package associates every power series with its
so-called {\em analytic expression}.  For each power series {\tt s},
created by the command {\tt PowerSeries} as the image of a polynomial
{\tt p} (under the natural embedding from ${\C}[X_1,\ldots,X_n]$ to
${\C}[[X_1,\ldots,X_n]]$) the polynomial {\tt p} is the analytic
expression of {\tt s}.  If a power series is defined by the sequence of
its homogeneous parts, as illustrated on
Figure~\ref{fig:PowerSeries-3}, the user can optionally specify the
{\em sum} of that series which is then set to its analytic expression.
Power series that have an analytic expression are closed under
addition, multiplication and inversion.  Propagating that information
provides the opportunity to speed up some computations and make
decisions that could not be made otherwise. For instance, the command
{\tt HenselFactorize} needs to decide whether its input polynomial has
an invertible leading coefficient; to do it starts by checking whether
the analytic expression of that leading coefficient is known and equal
to one.

The commands {\tt Display}, {\tt SetDefaultDisplayStyle} and {\tt
  SetDisplayStyle} control the output format of multivariate power
series and UPoPS.
Meanwhile, the commands {\tt HomogeneousPart}, {\tt Truncate},
{\tt GetCoefficient},  {\tt Precision}, {\tt Degree},
{\tt MainVariable} access data from a
power series or a univariate polynomial over power series,
as illustrated by Figure \ref{fig:PowerSeries-2}.

The commands {\tt Add}, {\tt Negate}, {\tt Multiply}, {\tt Exponentiate},
{\tt Inverse}, {\tt Divide}, {\tt EvaluateAtOrigin}, and {\tt TaylorShift}
perform arithmetic operations on multivariate power series and
univariate polynomials over multivariate power series. The
functionality of the first six commands can also be accessed using the
standard arithmetic operators. 
As will be discussed in Sections~\ref{sec:princ} and \ref{sec:impl}, the
implementation of every arithmetic operation, such as addition,
multiplication, inversion builds the resulting power series (sum,
product or inverse) ``lazily'', by creating its generator from the generators of
the operands, which are called {\em ancestors} of the resulting power
series.

\begin{figure}[H]
	\includegraphics[width=\textwidth]{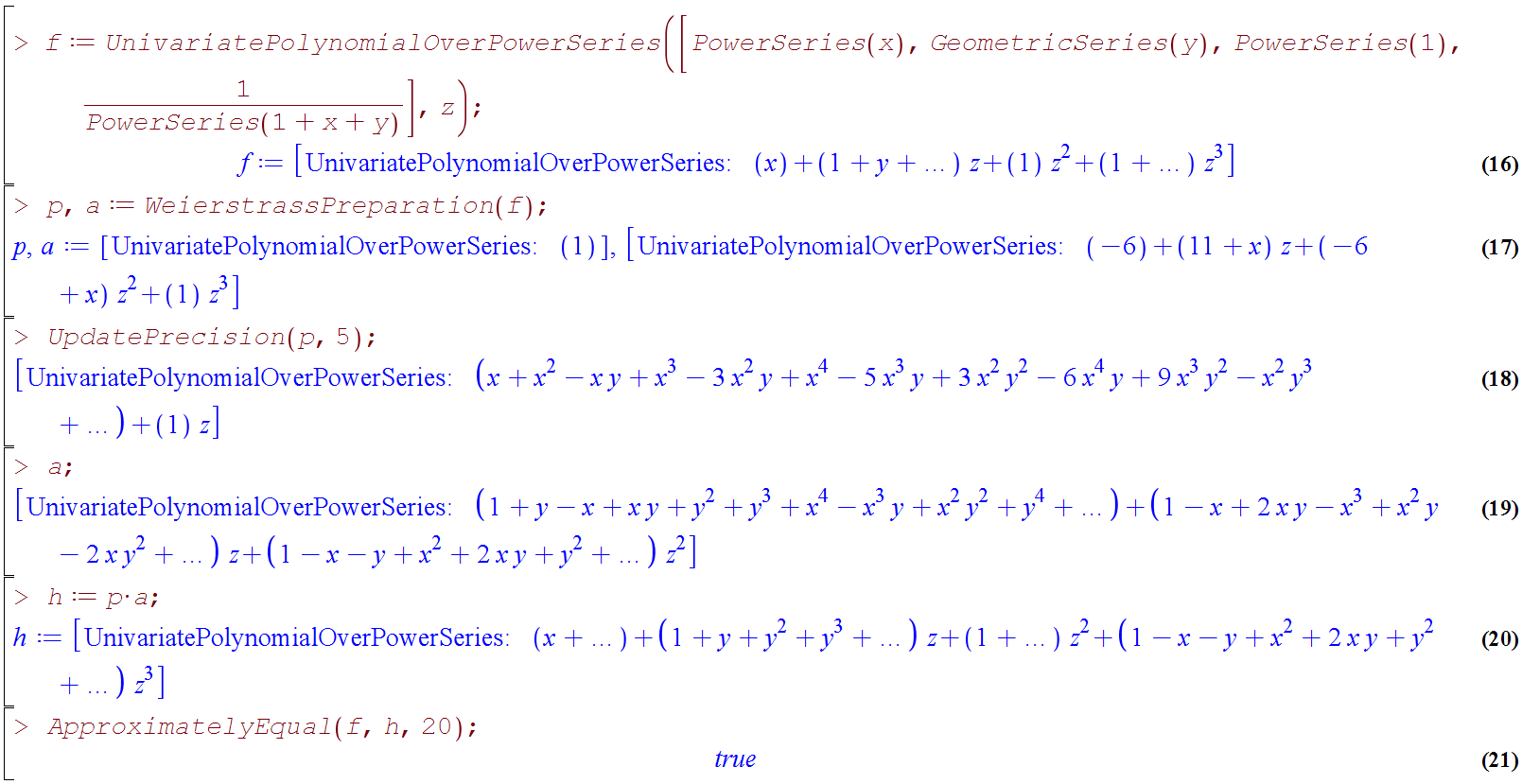}
	\caption{Factoring univariate polynomials using {\tt WeierstrassPreparation}.}
	\label{fig:PowerSeries-4}
\end{figure}

\begin{figure}[H]
	\centering
	\includegraphics[width=\textwidth]{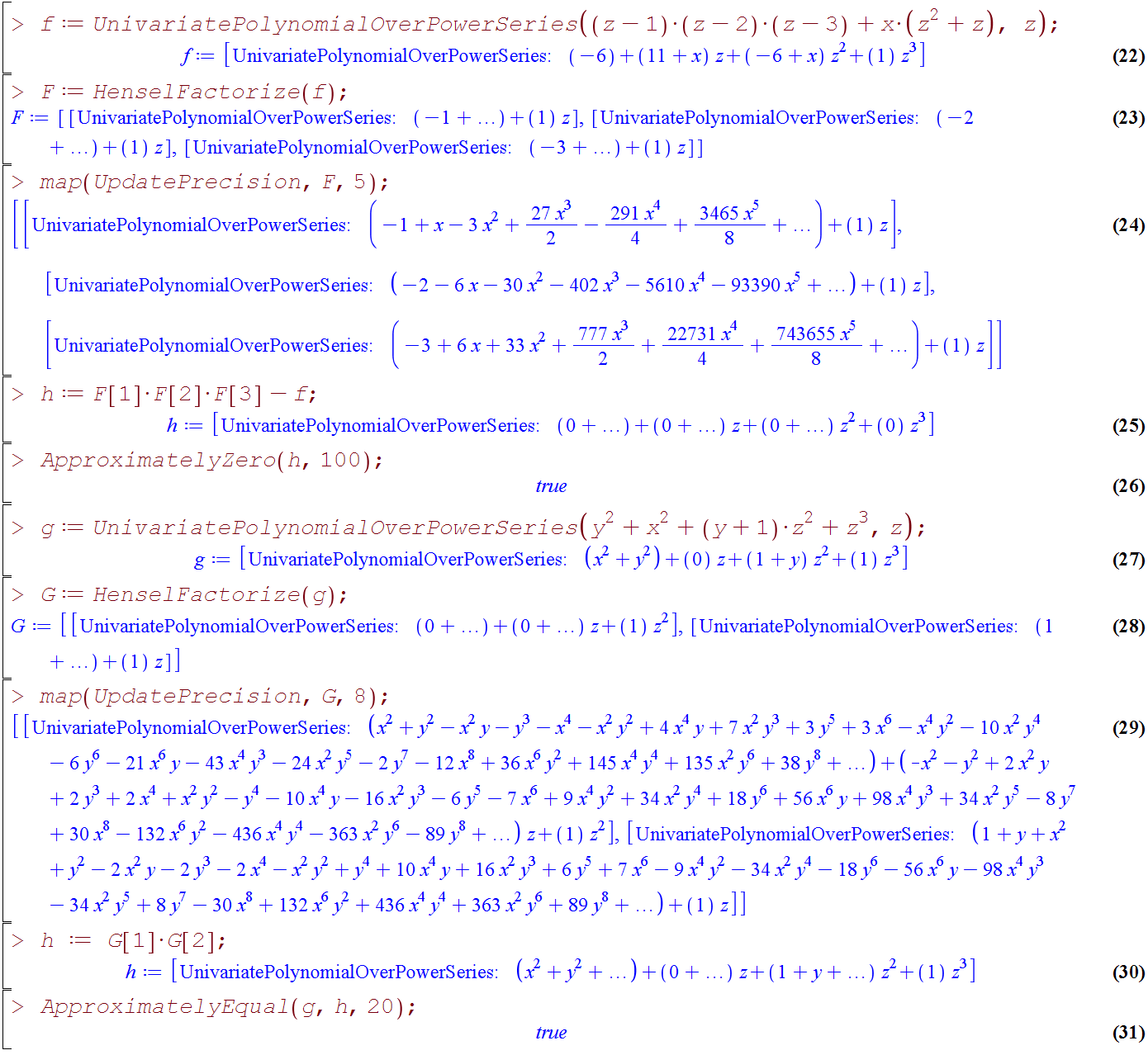}
	\caption{Factoring univariate polynomials using {\tt HenselFactorize}.}
	\label{fig:PowerSeries-5}
\end{figure}

\vspace{0.5cm}

The commands {\tt WeierstrassPreparation} and {\tt HenselFactorize} factorize
univariate polynomials over multivariate power series.
Thanks to their implementation based on lazy evaluation,
each of these factorization commands	
returns the factors as soon as enough information is discovered
for initializing the data structures of the factors;
see Figures~\ref{fig:PowerSeries-4} and \ref{fig:PowerSeries-5}.

The precision of each returned factor, that is, the common precision
of its coefficients (which are power series) is zero. However the
generator (see Section~\ref{sec:princ} for this term) of each
coefficient is known and, thus, the computation of more coefficients
can be resumed when a higher precision is requested. Such a request
can be explicit by calling \texttt{UpdatePrecision}, or implicit, 
when requesting data of a higher precision than has been previously 
requested through, e.g., \texttt{Truncate} or \texttt{HomogeneousPart}.

\newpage
\section{Design Principles}
\label{sec:princ}

In this section we examine several design principles 
underpinning the implementation of the {\MPS} library.
Foremost is lazy evaluation: an algorithmic technique 
where the computation of data is postponed until explicitly required (Section~\ref{sec:lazyeval}). 
The eventual implementations of these lazy-evaluation algorithms 
make deliberate efforts to use appropriate {\maple} data structures
and built-in functions to optimize performance (Section~\ref{sec:mapledatastruct}). 
Lastly, in support of software quality and integration with existing {\Maple} library code, 
we employ {\maple}'s object-oriented mechanisms (Section~\ref{sec:mapleobjects}).
\subsection{Lazy Evaluation}
\label{sec:lazyeval}
Lazy evaluation is an optimization technique most commonly appearing 
in the study of functional programming languages \cite{harper_2016}.
The lazy evaluation or ``call-by-need'' refers to delaying the
call to a function until its result is genuinely 
needed. This is often complemented by storing 
the result for later look-up.

In the case of power series, consider 
a bivariate geometric series 
$f = \sum_{d=0}^{\infty} f_{(d)}$
where $f_{(0)} = 1$, $f_{(1)} = x + y$, $f_{(2)} = x^2 + 2xy + y^2$,
\ldots, $f_{(d)} = (x + y)^d$.	
One can prove that $f$ converges to $\frac{1}{1-x-y}$.
\iffalse
In practice, it is impossible to store 
$f_{(d)}$ for all $d \geq 0$. 
\else
Of course, in practice, it is impossible to store
an infinite number of terms on a computer with finite memory. 
\fi
A na\"ive implementation then suggests storing $\Truncate{f}{d}$ for 
some large and predetermined $d$.
Thus, one can approximate power series as multivariate polynomials. 
Such an implementation could be called \textit{truncated power series}.
While this representation of power series is easy to implement, 
it leads to notable restrictions for the study of formal power series. 
\iffalse
First, one must determine the {\em precision}, i.e. the particular value of $d$, 
before initializing a power series and doing arithmetic as a priori. 
\else
First, one must a priori determine the {\em precision}, i.e. the particular value of $d$.
\fi
Second, in a most na\"ive implementation, 
previously-computed homogeneous parts must be recomputed
whenever a new, greater precision is required.
For example, the polynomial $f^{(d+1)}$ is likely to
be constructed ``from scratch'' despite the polynomial $f^{(d)}$ possibly
being already computed.
Third, storing and manipulating the polynomial part of a power series up 
to a degree $d$ needs a large portion of memory. 
This latter problem is exacerbated when the predetermined precision 
is not a tight upper bound on the required precision.

To combat the challenges of a truncated power series implementation, 
we take advantage of lazy evaluation.
Every power series is represented by a unique procedure 
to compute a homogeneous part for a given degree. 
For example, Listing \ref{lst:biGeoFunc} shows such
a procedure for the bivariate geometric 
series which converges to $\frac{1}{1-x-y}$. 
As we will see, 
this lazy evaluation 
design can be paired with an array of polynomials 
storing the previously computed 
homogeneous parts.

\begin{center}
\vspace{0.2em}
\begin{minipage}{\textwidth}
\begin{lstlisting}[language = Maple, caption = {A {\Maple} implementation of $f_{(d)}$ in $ \frac{1}{1-x-y} = \sum_{d=0}^{\infty} f_{(d)}$.}, label={lst:biGeoFunc}]
generator := proc(d :: nonnegint)
    return expand((x+y)^d); 
end proc;
\end{lstlisting}
\end{minipage}
\end{center}


\subsection{{\maple} Data Structures and Built-in Functions}
\label{sec:mapledatastruct}

Using an appropriate data structure for encoding 
and manipulating data is critical for performance, 
particularly in high-level and interpreted programming languages like {\Maple}.
In {\Maple}, modifying an existing list or
set---such as by appending, replacing, or deleting an element---leads 
to the creation of a new list or set, rather than modifying 
the original one in-place. 
\iffalse
Thus, we take advantage of {\tt Array}s, low-level mutable 
data-structures providing high-performance functionalities to store and 
update the allocation size and elements efficiently. 
\else
In contrast, an {\tt Array} is a low-level and mutable 
data-structure which allows for in-place modification of its elements. 
These functionalities provide much better performance than lists or sets
when the collection is frequently changed or when the elements being
modified are themselves large in size.
\fixed{This fact is clear from the overwhelming improvement
in performance of our library compared against the existing {\tt PowerSeries} library 
which uses lists to encode homogeneous parts; see Section~\ref{sec:exper}.}{I think this is true.. someone please confirm.}{Marc: yes, this is one of the reasons but not the only one.}
\fi

\iffalse
A deeper look in {\tt Array}s, 
\else
Looking more closely at the {\tt Array} data structure, 
\fi
an $n$-dimensional {\tt Array} is stored as a $n$-dimensional
rectangular block named {\tt RTABLE}. 
The length of the associated {\tt RTABLE} is $2n+d$ 
\iffalse
where $d$ is the number of elements; 
\else
where $d$ is maximum number of elements that may be stored,
i.e., the allocation size of the {\tt Array};
\fi
see \cite[Appendix 1]{mapleProgGuide}.
\iffalse
We utilize 1-dimensional arrays to store 
computed homogeneous parts named {\tt hpoly}, and 
coefficients of a UPoPS named {\tt upoly} in Listing \ref{lst:MPS}, so that,
the data fields point to multivariate polynomials and power series objects 
in {\tt hpoly} and {\tt upoly} respectively. 
\else
For the storage of homogeneous parts of a power series, 
and the power series coefficients of a UPoPS, 
we utilize 1-dimensional {\tt Array}s. 
Listing~\ref{lst:MPS} in the next section shows
this as the variables {\tt hpoly} and {\tt upoly}, respectively.
\fi

\iffalse
To provide a highly optimized library for formal power series, 
we follow advanced programming techniques and tricks.
\else
To further improve performance, we make use of  
low-level built-in functions. Such functions 
are provided as compiled code within the {\Maple} kernel, 
and therefore not written in the {\Maple} language.
\fi
Most notably, instead of using {\Maple} for-loops and the typical {\tt +} and {\tt *} syntaxes
for addition and multiplication, respectively,
we reduce the cost of summations and multiplications 
remarkably by taking advantage of built-in {\Maple} functions, {\tt add} and {\tt mul}.
These built-in functions, respectively, 
add or multiply the terms of an entire sequence of expressions together to return a single sum or product.
These functions avoid a large number of high-level function calls
and reduce memory usage by avoiding copying and re-allocation of data.
\removed{Moreover, we precisely make use of expansion in the entire library
as expanding polynomials 
is an expensive ubiquitous operation.}{I don't know what this is trying to say. Do you mean to say that expansion is avoided?}{There is the beginning of an idea here, but this would require more details in order to make sense. I think that it is obvious that a graded representation of power series requires to keep things expanded anyway.}

\subsection{{\maple} Objects}
\label{sec:mapleobjects}

\iffalse
A fundamental tool in \Maple's object-oriented programming 
mechanisms is Object, that allows 
variables and procedures to be encapsulated together. 
A power series object includes variables (object's members) 
and arithmetic procedures (object's methods). 
Objects can also restrict access to certain members and methods, and
members may be declares as {\it local} (or {\it export}) as they 
can be accessed from within the object's methods (or from everywhere). 
A class defines as a collection of all members and methods 
under a name, and objects are, in fact, instances of a class. 
By default, members and methods are unique within an object, 
unless they are declared as {\it static}. 
Then, they will be shared among all objects of a class. 
A power series class allows to overload basic operators, 
develop advanced display methods, 
create and manipulate power series objects, and 
write {\Maple} programs and interfaces on top of it; see \cite[Chapter 9]{mapleProgGuide} for more details about the object oriented programming in {\Maple}. 

We redesign data structures and algorithms of the power series 
and UPoPS object oriented. 
The classes for each are named, respectively,
\begin{itemize}
	\item[$\bullet$] {\PSO}, and
	\item[$\bullet$] {\UPoPSO}.
\end{itemize}
\else
An often overlooked aspect of {\maple} is its object-oriented capability.
An object allows for variables and procedures operating on that data 
to be encapsulated together in a single entity.
In {\maple}, a class---the definition of a particular type of object---can be
declared by including the option \texttt{object}
in a module declaration. 
Evaluating this declaration returns an object of that class. 
This new object is often a so-called ``prototype'' object which, 
when passed to the \texttt{Object} routine, returns a new object of the same class.
See \cite[Chapter 9]{mapleProgGuide} for further details on object-oriented programming in {\Maple}.

Our power series and UPoPS types are implemented using these
object-oriented features of {\Maple}.
The classes for each are named, respectively, 
{\PSO} and {\UPoPSO}.

The use of object-oriented programming in {\Maple} has two key benefits:
($i$) the organization object-oriented code provides better software quality through modularity and maintainability;
and ($ii$) allows for the overloading of built-in functions, thus allowing
objects to be integrated with, and used natively by, existing {\Maple} library functions.

\fi
\begin{lstlisting}[language = Maple, caption = {An overview of the {\MPS} package.}, label={lst:MPS}]
MultivariatePowerSeries := module()
option package;
	local PowerSeriesObject, 
		    UnivariatePolynomialOverPowerSeriesObject;
	# create a power series:
	export PowerSeries := proc(...) 
	# create a UPoPS:
	export UnivariatePolynomialOverPowerSeries := proc(...)
	#Additional procedures to interface these two classes
	
	module PowerSeriesObject()
	option object;
		local hpoly :: Array,
				  precision :: nonnegint,
				  generator :: procedure;
		# other members and methods 
	end module;	
	
	module UnivariatePolynomialOverPowerSeriesObject()
	option object;
		local upoly :: Array, vname :: name;
		# other members and methods 
	end module;
end module;
\end{lstlisting}

\iffalse
Both are integrated as local classes (modules) in the
{\MPS} \Maple~ library.  
Listing \ref{lst:MPS} shows an overview of this
which is developed by utilizing {\tt option package} \cite[Chapter 8]{mapleProgGuide}. 
This is in fact an interface package providing 
methods to construct, manipulate, and deconstruct 
objects from these two classes which will be discussed in Section \ref{sec:impl}.
\else
The {\MPS} library contains a package of the same name
which groups together those two aforementioned classes
along with additional procedures to construct and manipulate 
objects of those classes.
These additional procedures are used to ``hide'' 
the object-oriented nature of the library behind simple procedure 
calls.
This keeps the package syntactically and semantically
consistent with the general paradigm of {\Maple} which
does not use object-oriented programming.
As an example of such a procedure, \texttt{PowerSeries}, as seen
in Fig.~\ref{fig:PowerSeries-1} (Section~\ref{sec:UI}),
handles various different types of input parameters to correctly construct
a \texttt{PowerSeriesObject} object through delegation to the correct class method.

Listing~\ref{lst:MPS} shows the declaration of our
two classes and the {\MPS} package.
The latter is created by using {\tt option package} in a \texttt{module} declaration;
see \cite[Chapter 8]{mapleProgGuide}. 
The implementation of these two classes
is further discussed in Section~\ref{sec:impl}.
\fi

	\section{Implementation of {\MPS}}
\label{sec:impl}

The {\MPS} package
provides a collection of procedures 
which form simple wrappers for the methods 
of the aforementioned
classes, {\PSO} and {\UPoPSO}.\\
These classes, respectively, define the data structures and
algebraic functionalities for creating and manipulating 
multivariate power series and univariate polynomials
over power series. 
This section discusses those data structures
as well as the implementation of basic arithmetic, 
Weierstrass Preparation Theorem, and factorization 
via Hensel's lemma, all following a lazy evaluation scheme.
\subsection{\PSO}
\label{sec:PSO}
The {\PSO} class provides basic arithmetic operations, like
addition, multiplication, inversion, and evaluation, 
for multivariate power series, all utilizing lazy evaluation techniques. 
Let $f \in \KXN$ be a non-zero multivariate power series defined as 
$f = \sum_{d=0}^{\infty} f_{(d)}$. $f$ is encoded as an object of type {\PSO}, 
containing the following attributes. 
First, the power series 
{\em generator} is the procedure
to compute  $f_{(d)}$, the $d$-th homogeneous part of $f$,
for $d \in \N$.
Second, the {\em precision}
is a non-negative integer encoding
the maximum degree of the homogeneous parts 
which have so far been computed.
Third, the 1-dimensional array storing 
the previously computed homogeneous parts of $f$, denoted as 
{\tt hpoly} in Listing \ref{lst:MPS}.
To create a power series object this class provides a variety of constructors.
Power series objects may be created from polynomials,
algebraic numbers, UPoPS objects, or 
procedures defining the generator of the power series.

Every arithmetic operation returns a lazily-constructed
power series object by creating its generator from 
the generators of the operands,
but without explicitly computing any homogeneous parts of the result.
Thus, this is a lazy power series, so that, the homogeneous parts
of the result are computed when truly needed.
Once homogeneous parts are eventually computed,
they are stored in the array \texttt{hpoly}.
An important aspect of this organization 
is that the generator of the resulting power series becomes 
implicitly connected to the generators of the operands;
the latter are thus called the \textit{ancestors} of
the former.
\fixed{Note that the ancestors are merely stored as references, not copies, 
thus saving time and memory resources.
}{MA: check the correctness, please?}{I think we can just be vague and say ``references, not copies''} 


Moreover, the addition and multiplication operations 
are not only binary operations (operations taking two parameters), 
but are $m$-ary operations.
For multiplication, a sequence of power series $f_1, \ldots, f_m  \in \KXN$ may be passed to the multiplication algorithm 
to produce the product 
$f_1 \cdot f_2 \cdots f_m$ via lazy evaluation.
Similarly, addition may take the sequence $f_1, \ldots, f_m$ to
return the sum $f_1 + f_2 + \cdots + f_m$. 
Further, addition may also take as a parameter 
an optional sequence of polynomial coefficients $c_1, \ldots, c_m \in \K[X_1, \ldots, X_n]$ 
to return the sum $c_1 f_1 + \cdots + c_m f_m$ constructed lazily.
A key part to the efficiency of lazy evaluation is to not re-compute 
any data. We have already seen that the \texttt{hpoly} array
stores previously computed homogeneous parts for a {\PSO} object.
What is missing is to ensure that the array is accessed
where possible rather than calling the generator function.
Moreover, one must avoid directly accessing that array for homogeneous parts which are 
not yet computed.
We thus provide the function {\tt HomogeneousPart}$(f, d)$, demonstrated in Listing \ref{lst:hpart},
to handle both of these cases.
This function returns the $d$-th homogeneous part of the power series $f$;
if $d$ is greater than the {\em precision} ({\tt f:-precision}), then this method iteratively calls the {\em generator}
to update {\tt hpoly} and {\tt precision}, otherwise it simply returns the previously computed homogeneous part.
From here on we use \texttt{hpart} as shorthand for the \texttt{HomogeneousPart} function.
\begin{lstlisting}[mathescape = true, language = Maple, caption = {A simplified version of the {\tt HomogeneousPart} function in {\PSO}.}, label={lst:hpart}]
export HomogeneousPart ::static := proc(f, d :: nonnegint)
	if d > f:-precision then
		f:-hpoly(d+1) := 0; # resize the hpoly array
		for local i from f:-precision + 1 to d do 
			f:-hpoly[i] := f:-generator[i];
		end do;
		f:-precision := d;
	end if;
	return f:-hpoly[d];
end proc;
\end{lstlisting}
Listing \ref{lst:div} shows a simplified implementation of 
{\tt Divide} that computes the quotient of two
power series objects $f, g \in \KXN$.  
In particular, notice the creation of the
local procedure \texttt{gen} for the generator of the quotient.
Note that {\tt EXPAND} is a local macro defined in {\MPS}
to efficiently perform expansion and normalization 
supporting {\em algebraic} inputs.
\begin{lstlisting}[mathescape = true, language = Maple, caption = {A simplified version of the division method in {\PSO}.}, label={lst:div}]
export Divide ::static := proc(f, g)
	if hpart(g,0)=0 then 
		error "invalid input: not invertible"; 
	end if;
	local h := Array(0..0,EXPAND(hpart(f,0)/hpart(g,0)));
	local gen := proc(d :: nonnegint)
		local s := hpart(f,d);
		s -= add(EXPAND(hpart(g,i)*hpart(f,d-i)),i=1..d);
		return EXPAND(s/hpart(g,0));
	end proc;
	return Object(PowerSeriesObject,h,0,gen);
end proc;
\end{lstlisting}
\subsection{\UPoPSO}
\label{sec:upopso}

The {\UPoPSO} class is implemented as a simple dense univariate polynomial with the simple and obvious implementations
of associated arithmetic (see, e.g., \cite[Chapter 2]{Gathen:2013:MCA}).
The arithmetic operations are achieved directly from coefficient arithmetic, that is, {\PSO} arithmetic.
Since the latter is implemented using lazy evaluation techniques, 
UPoPS arithmetic is inherently and automatically lazy.
For example, the addition of two UPoPS objects
$f = \sum_{i = 0}^{k} a_i {X_{n+1}^i}$ and $g = \sum_{i=0}^{k} b_i {X_{n+1}^i}$ in $\KXN[{X_{n+1}}]$
is the summation $(a_i + b_i){X_{n+1}^i}$ for all $ 0 \leq i \leq k$, where $a_i, b_i$ are {\PSO} objects.
Other basic arithmetic operations behave similarly.
However, there are important operations on UPoPS which are not as straightforward. 
In the following we explain our implementation of Weierstrass Preparation Theorem, Taylor shift, 
and factorization via Hensel's lemma for UPoPS, all of which follow lazy evaluation techniques.
\medskip\noindent{ \bf Weierstrass Preparation}.
Let $f, p, \alpha \in \KXN[{X_{n+1}}]$ be such that they satisfy the conditions of Theorem~\ref{WeierstrassTheorem} and
such that $f = \sum_{i = 0}^{d + m} a_i {X_{n+1}^i}$, $p = {X_{n+1}}^d + \sum_{i=0}^{d-1} b_i {X_{n+1}^i}$, and $\alpha = \sum_{i=0}^{m} c_i {X_{n+1}^i}$.
Equating coefficients in $f = p \alpha$ we derive the two following systems of equations: 
\begin{equation}
\label{eq:WPT-1}
\begin{cases} \begin{array}{lcl} 
{ a_0} & = & \  { b_0} { c_0} \\
{ a_1} & = & \ b_0 c_1 + {  b_1} { c_0} \\
& \vdots &  \\
{ a_{d-1}} & = & \ b_0 c_{d-1} + b_1 c_{d-2} + \cdots + b_{d-2} c_1 + { b_{d-1}} { c_0} 
\end{array} \end{cases} 
\end{equation}
\begin{equation}
\label{eq:WPT-2}
\hspace{-1.6em}
\begin{cases} \begin{array}{lcl} 
{  a_d} & = & \ b_0 c_d + b_1 c_{d-1} + \cdots + b_{d-1} c_1 + { c_0}  \\ 
& \vdots &  \\
{ a_{d+m-1}} & = & b_{d-1} c_{m} + { c_{m-1}} \\
{ a_{d+m}}  & = & { c_m} 
\end{array} \end{cases}
\end{equation}
To solve these systems we proceed by solving them modulo successive powers of $\M$, 
following the proof of Theorem~\ref{WeierstrassTheorem} in~\cite{brandt2020power}.
Notice that solving modulo successive powers of $\M$ 
is precisely the same as computing homogeneous parts of 
increasing degree. Thus, this follows our lazy evaluation scheme perfectly.
The power series $b_0, \ldots, b_{d-1}$ are generated by Equations~(\ref{eq:WPT-1})
and $c_0, \ldots, c_m$ by Equations~(\ref{eq:WPT-2}).
Consider that $b_0, \ldots, b_{d-1}, c_0, \ldots, c_{m}$ are known modulo
$\M^r$ while $a_0, \ldots, a_{d-1}$ are known modulo $\M^{r+1}$;
this latter fact is simple since $f$ is the input to Weierstrass Preparation and is fully known.
From the first equation in (\ref{eq:WPT-1}), $b_0$ can be computed modulo $\M^{r+1}$
since $b_0 \in \M$, $c_0$ is known modulo $\M^r$, and $a_0$ is known $\M^{r+1}$. 
Then, the equation $a_1 = b_0c_1 + b_1c_0$, that is,  $a_1 - b_0c_1 = b_1c_0$ can be solved for $b_1$ modulo $\M^{r+1}$ 
since, again, $b_1 \in \M$ and the other terms are sufficiently known.
We compute all $b_{2}, \ldots, b_{d-1}$ modulo $\M^{r+1}$ with the same argument.
After determining $b_0, \ldots, b_{d-1}$ modulo $\M^{r+1}$, 
we can compute $c_m, c_{m-1},\ldots, c_0$ modulo $\M^{r+1}$
from Equations~(\ref{eq:WPT-2}) with simple power series multiplication 
and subtraction, working iteratively, in a  bottom up fashion. 
For example, $c_{m-1} = a_{d+m-1} - b_{d-1}c_m$.

As yet, we have not explicitly seen how the coefficients of $p$ and $\alpha$ will be updated.
The key idea is that to update a single power series coefficient of $p$ or $\alpha$ 
requires simultaneously updating all coefficients of $p$ and $\alpha$.
Thus, all the generators of $b_0, \ldots, b_{d-1}, c_0, \ldots, c_m$ simply call
a single ``Weierstrass update'' function to update all power series simultaneously
using Equations~$(1)$ and $(2)$.
Algorithm \ref{alg:WPupdate} shows this Weierstrass update function. 

{
\setlength{\textfloatsep}{0.5em}
\setlength{\intextsep}{0.8em}
\begin{algorithm}[htb]
	\caption{
		\small {\sc WeierstrassUpdate}$(p, \alpha, \F, r)$\newline
		Given $p = {X_{n+1}^d} + \sum_{i=0}^{d-1} b_i {X_{n+1}^{i}}$, $\alpha = \sum_{i=0}^{m} c_i {X_{n+1}^i}$, $r \in \N$, and $\F = \{ F_i \ \vert \ F_i = a_i - \sum_{j=0}^{i-1} b_{j} c_{i-j}, 0 \leq i < d \}$ are all known modulo ${\M}^{r}$, returns $b_0, \ldots, b_{d-1}, c_0, \ldots, c_m$ modulo ${\M}^{r+1}$.} \label{alg:WPupdate}
	\algnotext{EndIf}
	\algnotext{EndFor}
	\begin{algorithmic}[1]
		\Statex {$\verb|#|$ update $b_0, ..., b_{d-1}$ modulo $\M^{r+1}$} 
		\For{$i$ \textbf{from} $0$ \textbf{to} $d-1$}
		\State $s := \text{\tt add}(\text{\tt seq}( {\tt hpart}(b_i, r-k) \cdot {\tt hpart}(c_{_0}, k), \ k =  1 \ .. \ r-1 ))$;
		\State ${\tt hpart}(b_i, r) := ({\tt hpart}(F_i, r) - s)/{\tt hpart}(c_0, 0)$;
		\EndFor
		\Statex {$\verb|#|$ ensure $c_0, ..., c_{m}$ are updated modulo $\M^{r+1}$} 
		\For{$i$ \textbf{from} $0$ \textbf{to} $m$}
		\State ${\tt hpart}(c_i, r)$; 
		\EndFor	
	\end{algorithmic}
\end{algorithm}
}

In order to update the coefficients of $p$, we frequently need to compute $a_i - \sum_{j=0}^{i-1} b_{j} c_{i-j}$ for $0 \leq i < d$.
To optimize this operation, we a priori create helper power series as the set
$\F = \{ F_i \ \vert \ F_i = a_i - \sum_{j=0}^{i-1} b_{j} c_{i-j}, i = 0, \ldots, d-1 \}$.
The power series $F_i$, following power series arithmetic with lazy evaluation, 
allows for the efficient computation of 
homogeneous parts of increasing degree of $a_i - \sum_{j=0}^{i-1} b_{j} c_{i-j}$.
This set $\F$ is passed to the Weierstrass update function to optimize the overall computation.

Finally, the Weierstrass preparation must be initialized before continuing with Weierstrass updates.
Namely, the degree of $p$ and the initial values
of $p$ and $\alpha$ modulo $\M$ must first be computed.
The degree of $p$, namely $d$, is set to be the smallest integer $i$ such that 
$a_i$ is a unit. If $d = 0$, then $p = 1$ and $\alpha = f$, otherwise,
$m$ equals the difference between the degree of $f$ and $d$, 
and we initialize $b_i = 0$ for $ 0 \leq i < d$.
Then, $c_m, \ldots, c_0$ are initialized using power series arithmetic following Equations~(\ref{eq:WPT-2}).
Lastly, the set $\F$ is initialized.
\medskip\noindent{ \bf Taylor Shift}.
This operation takes a UPoPS object $f \in \KXN[{X_{n+1}}]$
and performs the translation ${X_{n+1}} \rightarrow {X_{n+1}} + c$, i.e. $f({X_{n+1}}+c)$, for some $c \in \K$. 
In our implementation, $c$ can be a {\tt numeric} or {\tt algebraic} \Maple~type
with the purpose of being used efficiently in factorization via Hensel's Lemma. 

Assume $f = \sum_{i=0}^{k} a_i {X_{n+1}^i}$ is a UPoPS in $\KXN[{X_{n+1}}]$ and $c \in \K$.
As the {\PSO} objects $a_0, \ldots, a_k$ are lazily evaluated power series,
we want to also make Taylor shift  a lazy operation. 
Thus, we need to create a generator for the power series coefficients of $f({X_{n+1}}+c)$.
Let $\T = (t_{i,j})$ be the lower triangular matrix of the coefficients of ${X_{n+1}}^j$
in the binomial expansion $({X_{n+1}}+c)^i$, for $0 \leq i \leq k$, and $0 \leq j \leq i$.
Let $\left( b_0, \ldots, b_{k} \right)$ be the list of coefficients of $f({X_{n+1}}+c)$ in {\KXN}.
Then, it is easy to prove that for every $0 \leq i \leq k$, 
$b_i$ is the inner product of the $i$-th sub-diagonal of $\T$
with the lower $k+1-i$ elements of the vector $\left( a_0, \ldots, a_k \right)$.
This inner product can be computed efficiently 
by taking advantage of the $m$-ary addition operation described for the {\PSO} (see Section~\ref{sec:PSO}).
Since this operation returns a lazily-constructed power series, 
this precisely defines the lazy construction of the power series $b_0, \ldots, b_k$, 
thus making Taylor shift a lazy operation.

\medskip\noindent{ \bf Factorization via Hensel's Lemma}.
Hensel's lemma for factorizing 
univariate polynomials over power series 
was reviewed in Theorem \ref{HenselLemma}, 
where $\K$ is algebraically closed and $f \in \KXN[{X_{n+1}}]$ is a UPoPS object. 
Following the ideas of~\cite{brandt2020power},
we compute the factors of $f$ in a lazy fashion.
Algorithm~\ref{alg:fact}  proceeds through iterative applications of 
Taylor shift and Weierstrass Preparation Theorem in order to create 
one factor of $f$ at a time. Those factors are actually computed
through lazy evaluation thanks to the lazy behavior of
the procedures {\sc WeierstrassPreparation} and {\sc  TaylorShift}.
This Algorithms thus computes and updates the factors modulo the successive powers ${\M}, {\M}^2, {\M}^3, \ldots$ 
of the maximal ideal ${\M}$.

\begin{algorithm}[bth]
	\caption{
		\small \HF$(f)$\newline
		Given $f = \sum_{i=0}^{k} a_i {X_{n+1}}^{i} \in \KXN[{X_{n+1}}]$, returns a list of factors $\{ f_1, \ldots, f_r \}$ so that $f = a_k \cdot f_1 \cdots f_r$, and satisfies Theorem \ref{HenselLemma}.} \label{alg:fact}
	\algnotext{EndIf}
	\algnotext{EndFor}
	\begin{algorithmic}[1]
		\If{$a_k \notin \M$}
			\State $f^{*} := \frac{1}{a_k} \cdot f$; 
		\Else 
			\State \textbf{error} ``$a_k$ must be a unit." 					
		\EndIf
		\State $\bar{f} := \EAO(f^{*})$;
		\State $c_1, \ldots, c_r := \text{\sc Roots}(\bar{f}, {X_{n+1}})$;
		\For{$i$ \textbf{from} $1$ \textbf{to} $r$}
		\State $g := \TS(f^*, c_i)$;
		\State $p, \alpha := \WP(g)$;
		\State $f_i := \TS(p, -c_i)$;
		\State $f^* := \TS(\alpha, -c_i)$;
		\EndFor
		\State \Return $\{ f_1, \ldots, f_r \}$;
	\end{algorithmic}
\end{algorithm}
Note that the generation of the factors $f_1, \ldots, f_r $
takes place after factorizing  $\bar{f} \in \K[{X_{n+1}}]$. 
Recall that $\bar{f}$ is obtained 
by evaluating each $X_i$ to $0$ for $1 \leq i \leq n$.
This is called {\EAO} in our implementation. 
To efficiently factor $\bar{f}$, we take advantage of
the package {\tt SolveTools}~\cite{mapleSolveTools},
which allows us to compute the splitting field of $\bar{f}$
(which, in practice, is a polynomial with coefficients
in some algebraic extension of {\Q})
and factorize $\bar{f}$ into linear factors.

Let $c_1, \ldots, c_r$ be 
the distinct roots of $\bar{f}$ 
and $k_1, \ldots, k_r$ their respective multiplicities.
To describe one iteration of  
Algorithm~\ref{alg:fact}, 
let $f^{*}$ be the current polynomial to factorize. 
For a root $c_i$ of $\bar{f}$, and thus $f^*$, 
we perform a Taylor shift to obtain $g = f^{*}({X_{n+1}} + c_i)$.
Then, we apply Weierstrass preparation 
on $g$ to obtain $p$ and $\alpha$ 
where $p$ is monic and of degree $k_i$. 
Again, by using Taylor Shift, 
we apply the reverse shift to $p$ to obtain $f_i = p({X_{n+1}} - c_i)$, 
a factor of $f$, 
and $f^{*} = \alpha({X_{n+1}} - c_i)$, 
for the next iteration. 
As mentioned above, since both Taylor shift and Weierstrass preparation
are implemented using lazy evaluation, our factorization via Hensel's lemma 
is inherently lazy. 
	\section{Experimentation}
\label{sec:exper}
We compare the performance of the {\MPS} package, denoted {\tt MPS}, with 
the previous {\Maple} implementation of multivariate power series, the \texttt{PowerSeries} package,
denoted {\tt RCPS}, and
the recent implementation of power series via lazy evaluation 
in the {\BPAS} library. 
This latter implementation is written in the C language on top 
of efficient sparse multivariate arithmetic; see \cite{asadiAlgorithms2019,brandt2020power}.
It has already been shown in \cite{brandt2020power} 
that the implementation in {\BPAS} is orders of magnitude
faster than the \texttt{PowerSeries} package, 
{\Maple}'s \texttt{mtaylor} command, and 
the multivariate power series available in {\Sage}.
As we will see, our implementation performs comparably to
that of {\BPAS}.

\begin{figure}[!b]
	\centering
	\begin{minipage}[b]{0.49\textwidth}
	    \includegraphics[width=\textwidth]{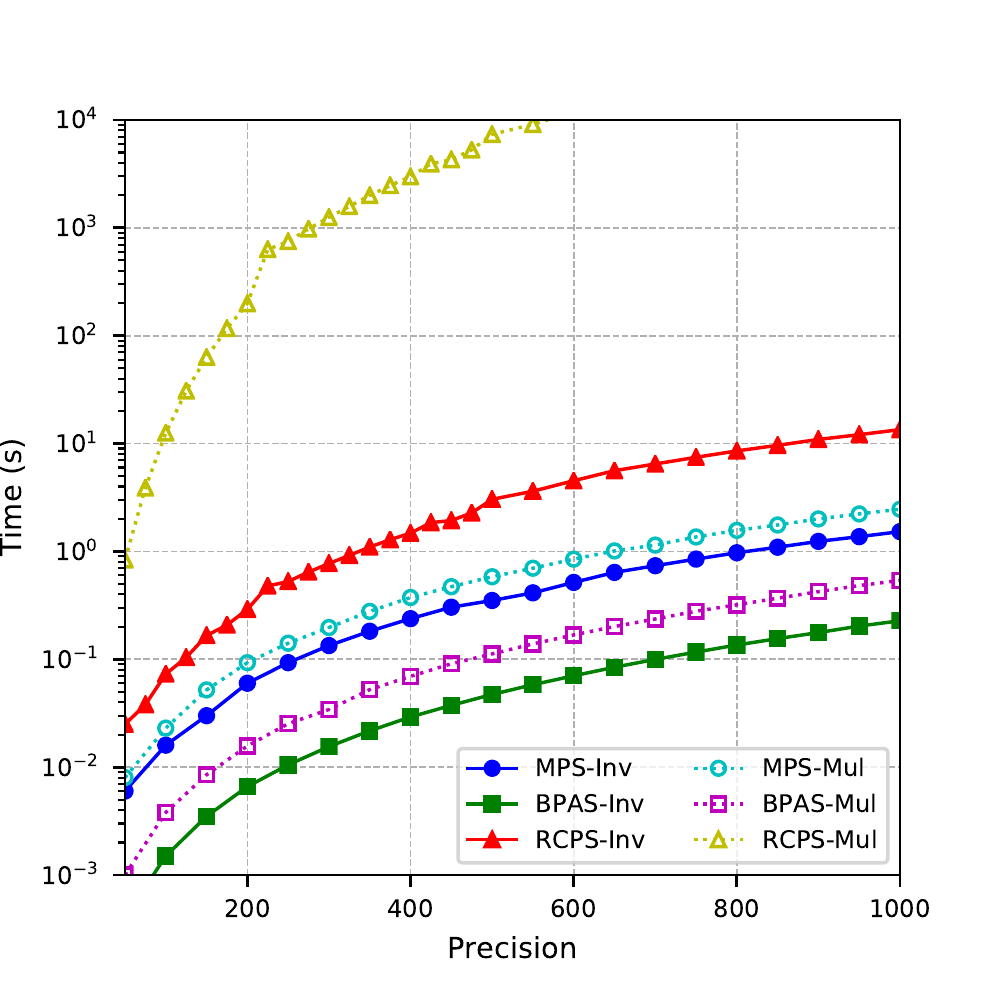}
		\caption{Computing $\frac{1}{f}$ and $\frac{1}{f} \cdot f$ for \newline $f_1 = 1 + X_1 + X_2$.} 
		\label{fig:inv-t1}
	\end{minipage}
	\begin{minipage}[b]{0.49\textwidth}
		\includegraphics[width=\textwidth]{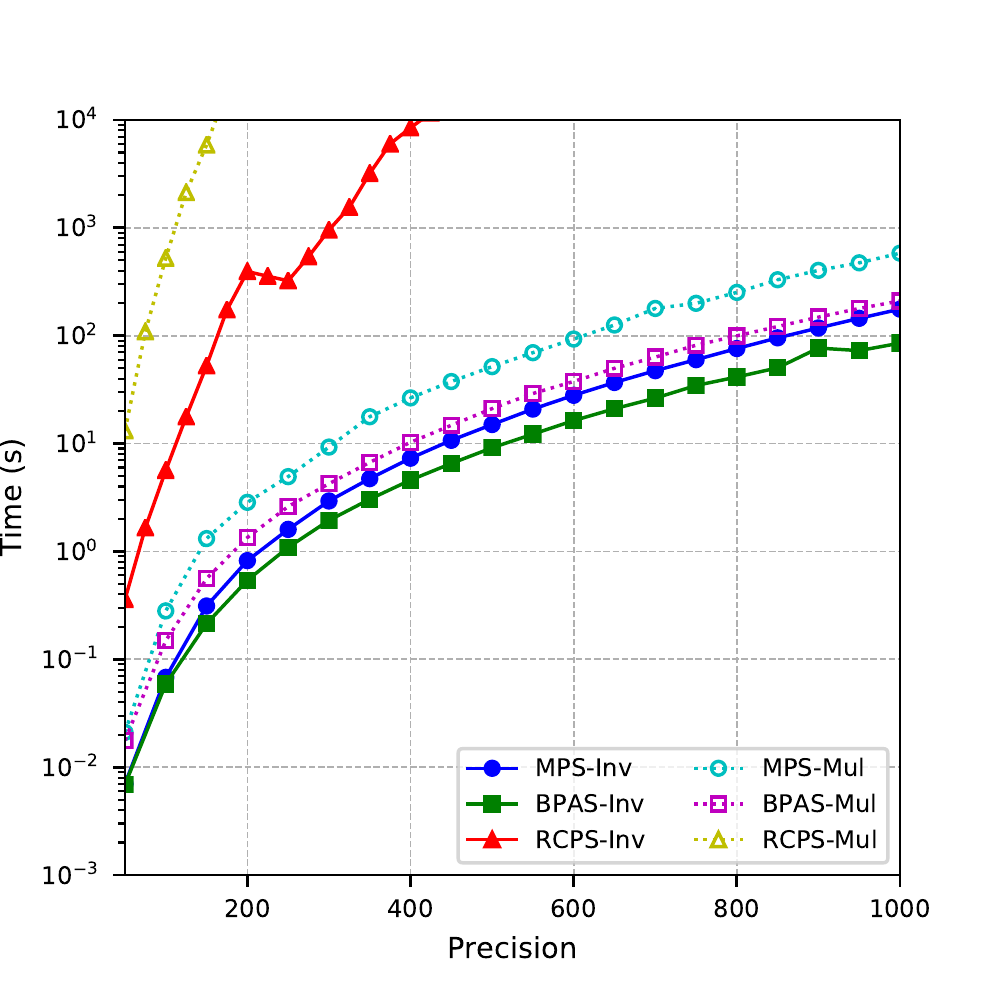}
		\caption{Computing $\frac{1}{f}$ and $\frac{1}{f} \cdot f$ for \newline $f_2 = 1 + X_1 + X_2 + X_3$.} 
		\label{fig:inv-t2}
	\end{minipage}
\end{figure}

\begin{figure}[!t]
	\centering
	\begin{minipage}[b]{0.49\textwidth}
		\includegraphics[width=\textwidth]{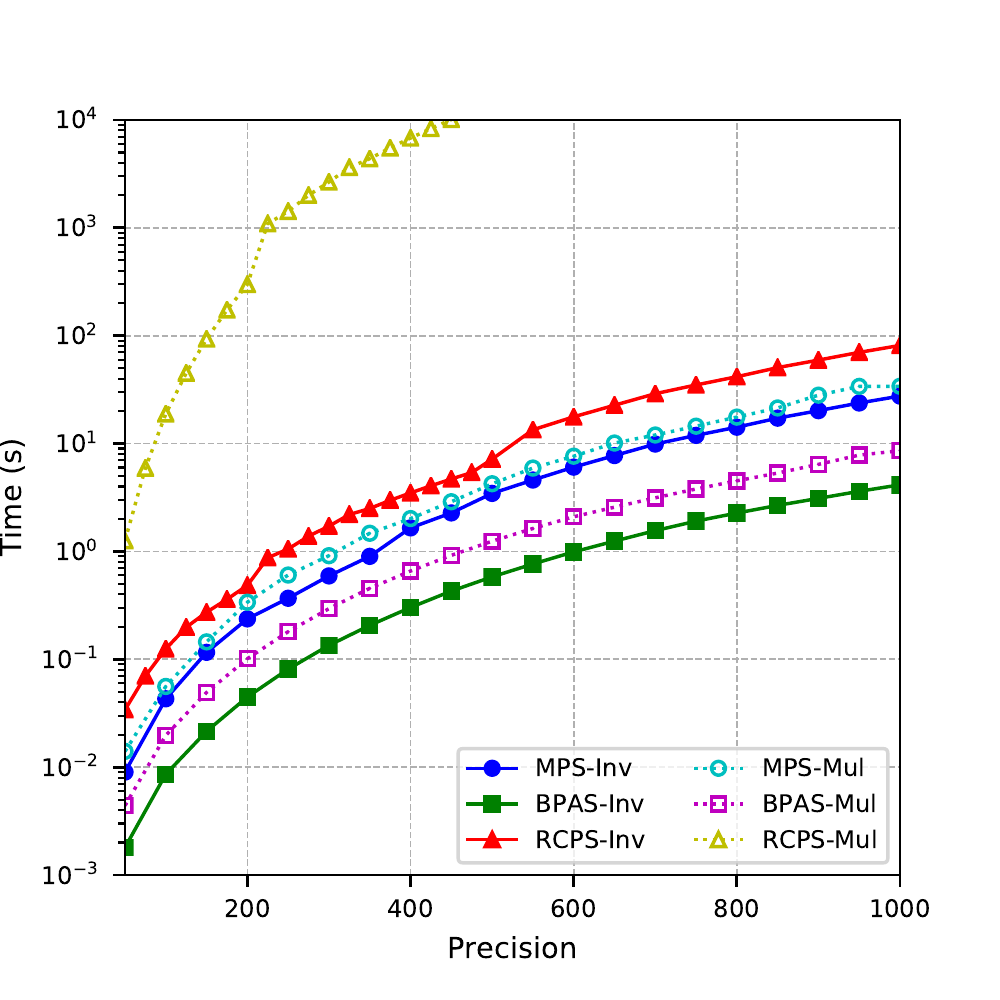}
		\caption{Computing $\frac{1}{f}$ and $\frac{1}{f} \cdot f$ for \newline $f_3 = 2 + \frac{1}{3}(X_1 + X_2)$.} 
		\label{fig:inv-t4}
	\end{minipage}
	\begin{minipage}[b]{0.49\textwidth}
	\includegraphics[width=\textwidth]{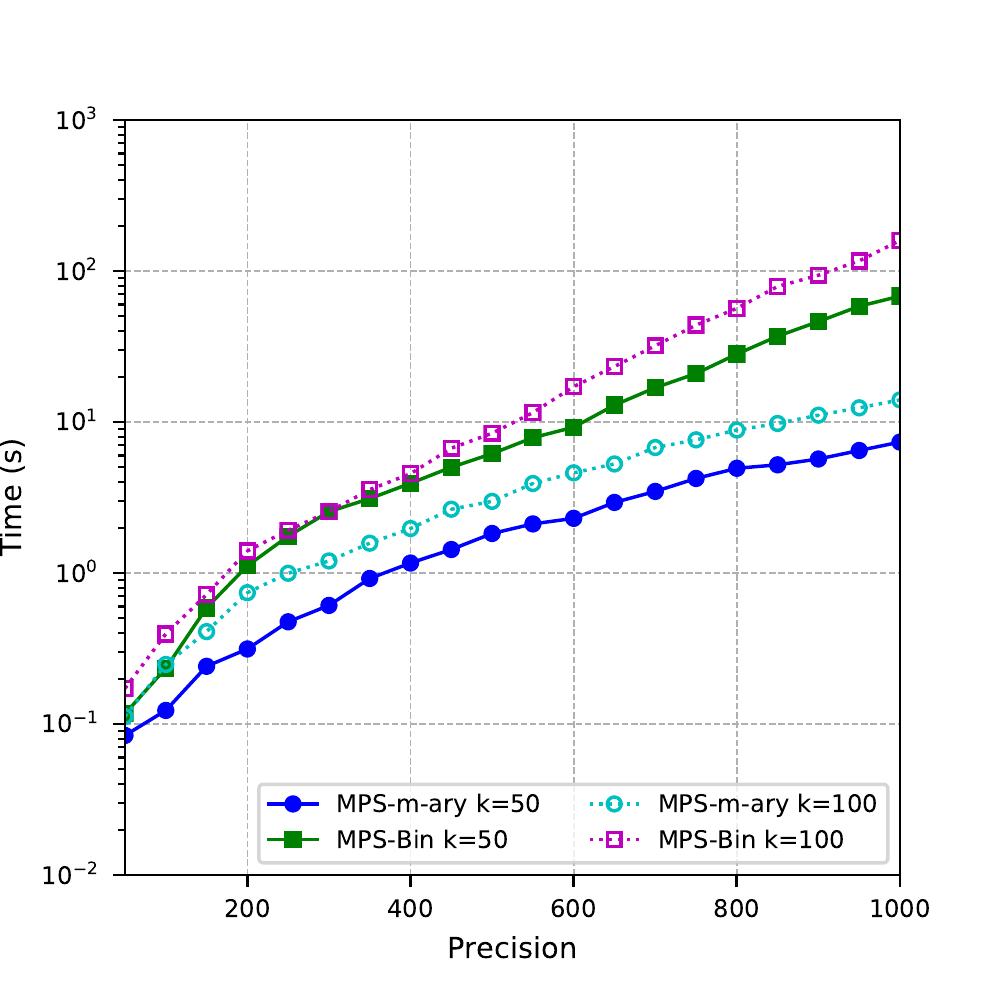}
	\caption{Computing $f = \sum_{i=1}^{k} \frac{1}{1-x-y}$ using $m$-ary and binary addition.}
	\label{fig:nary}
	\end{minipage}
\vspace{0.15em}
\end{figure}

Throughout this section, we collect our benchmarks
on a machine running Ubuntu 18.04.4, {\Maple} 2020, 
and {\BPAS} (ver. 1.652), with an Intel Xeon X5650 processor 
running at 2.67GHz, with 12x4GB DDR3 memory at 1.33 GHz. 
Figures \ref{fig:inv-t1}, \ref{fig:inv-t2}, and \ref{fig:inv-t4}, respectively, show
the performance of division and multiplication algorithms to compute
$\frac{1}{f}$ and $\frac{1}{f} \cdot f$ for power series $f_1 = 1 + X_1 + X_2$, 
$f_2 = 1 + X_1 + X_2 + X_3$, and $f_3 = 2 + \frac{1}{3}(X_1 + X_2)$. 
It can be seen that {\tt MPS} power series division is 9$\times$, 2100$\times$, and 3$\times$ 
faster than the previous {\Maple} implementation for $f_1, f_2$, and $f_3$ respectively.
The speed-ups for multiplication are significantly higher. Moreover, {\tt MPS} 
results are comparable with the C implementation of similar algorithms in {\BPAS}.
Figure~\ref{fig:nary} then highlights the efficiency 
of $m$-ary addition (see Section~\ref{sec:PSO}), compared to
iterative applications of binary addition.
Recall that $m$-ary addition is exploited in the Weierstrass preparation algorithm. 

\begin{figure}[!t]
	\centering
	\begin{minipage}[b]{0.49\textwidth}
		\includegraphics[width=\textwidth]{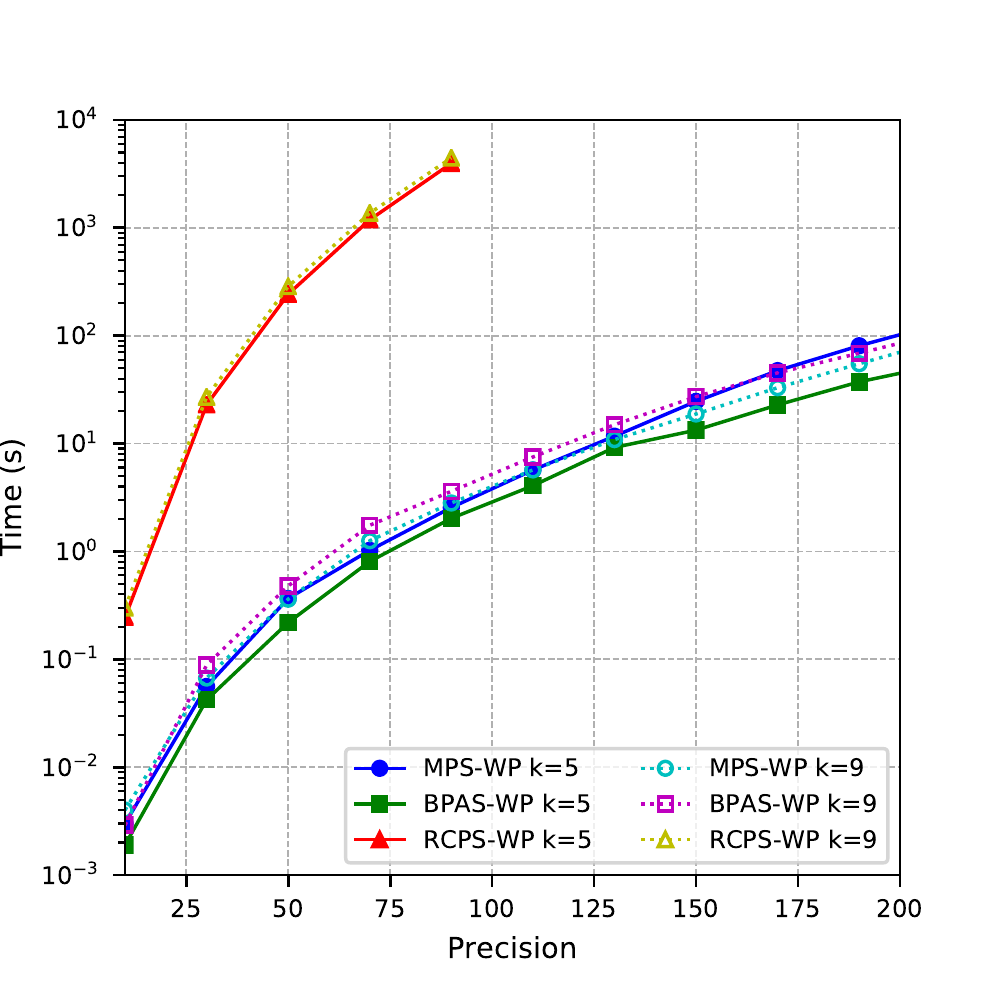}
		\caption{Computing Weierstrass preparation of {$f_1 = \frac{1}{1+X_1+X_2}{X_{3}}^{k} + {X_{3}}^{k-1}+\cdots+X_2 {X_{3}} + X_1 \in \K[\![X_1, X_2]\!][X_3]$}.}
		\label{fig:wp-t1}
	\end{minipage}
	\begin{minipage}[b]{0.49\textwidth}
		\includegraphics[width=\textwidth]{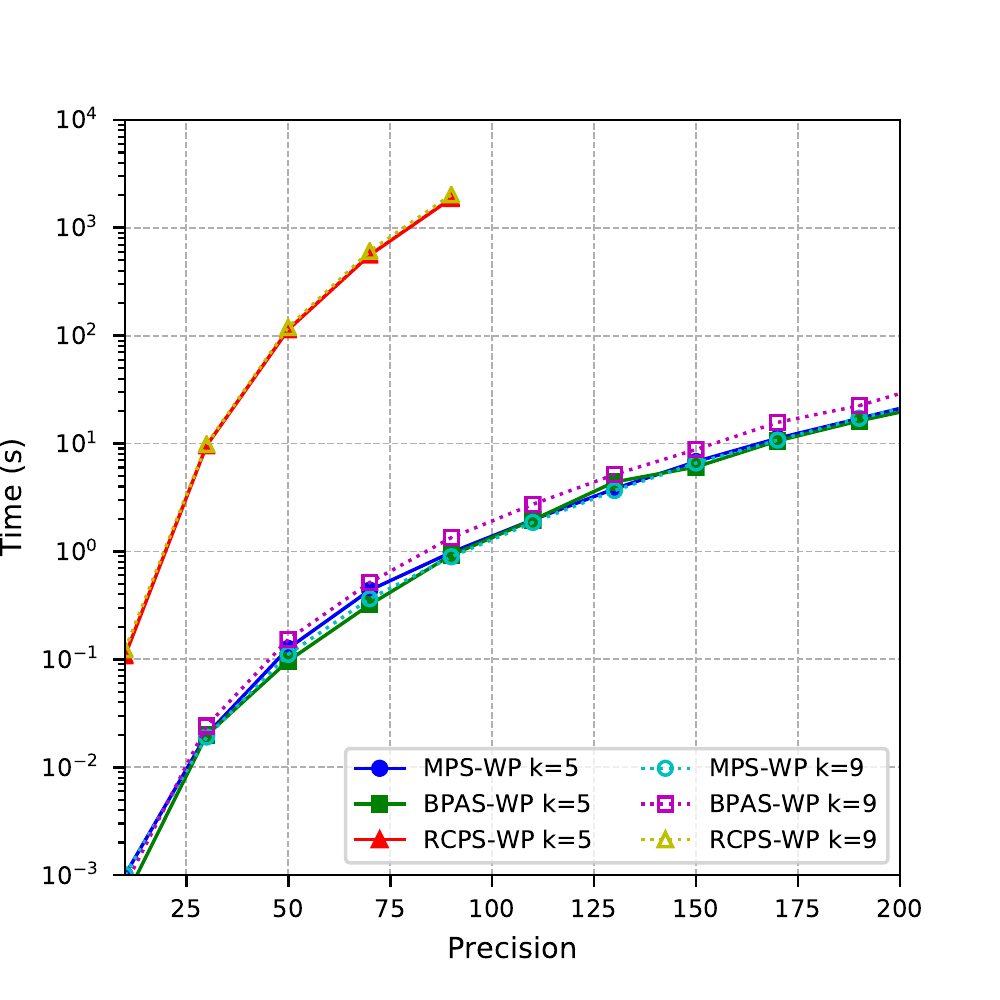}
		\caption{Computing {Weierstrass preparation} of {$f_2 = \frac{1}{1+X_1+X_2}{X_{3}}^{k} + X_2 {X_{3}}^{k-1}+\cdots+ {X_{3}} + X_1 \in \K[\![X_1, X_2]\!][X_3]$}.} 		\label{fig:wp-t2}
	\end{minipage}
\end{figure}

Next, we compare the performance of Weierstrass preparation (Section~\ref{sec:upopso}).
Figures \ref{fig:wp-t1} and \ref{fig:wp-t2} demonstrate 
the running time of this algorithm for two different UPoPS. 
Looking at these results, we can see a 2200$\times$ speed-up 
in comparison with the similar algorithm in {\tt RCPS}
and timings comparable to {\BPAS}.

\begin{figure}[!thp]
	\centering
	\begin{minipage}[b]{0.49\textwidth}
		\includegraphics[width=\textwidth]{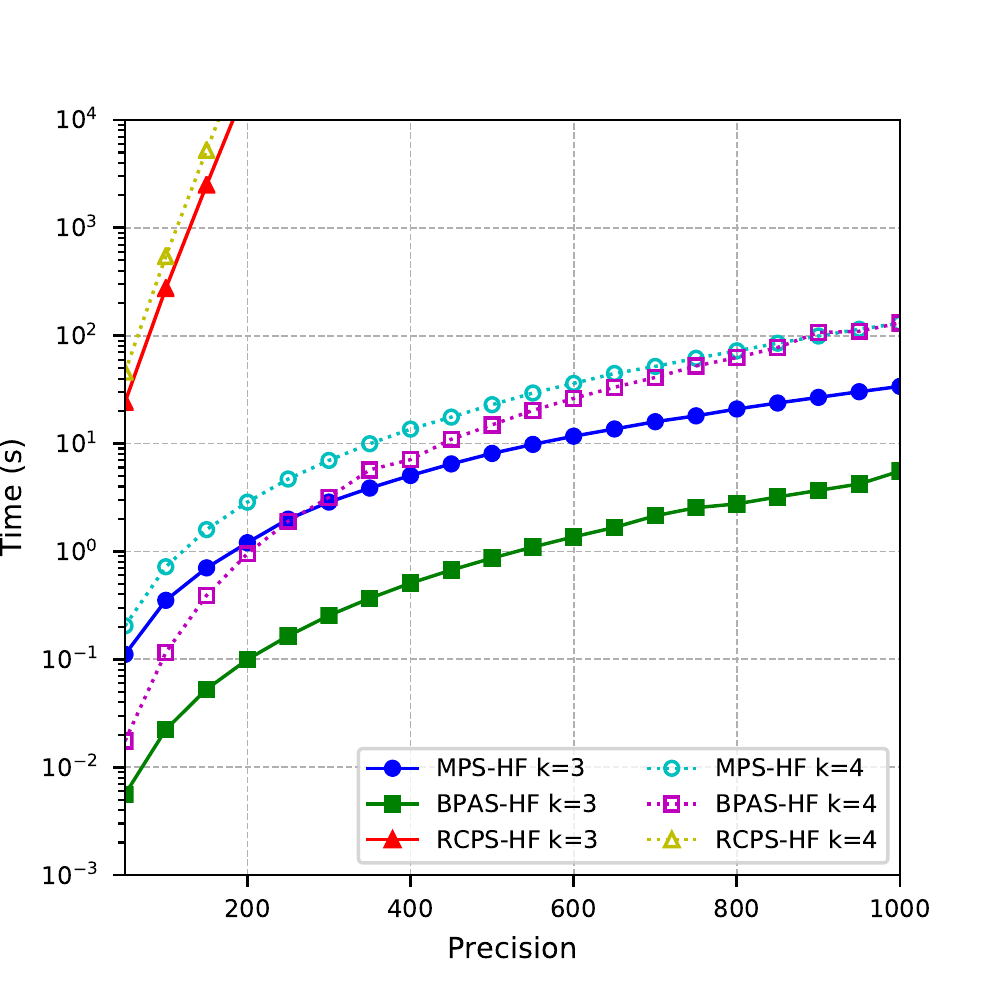}
		\caption{Computing ${\HF}(f)$ for { $f = \prod_{i=1}^{k}({X_{2}}-i) + X_1({X_{2}}^{k-1}+{X_{2}})$}.}
		\label{fig:fact}
	\end{minipage}
	\begin{minipage}[b]{0.49\textwidth}
		\includegraphics[width=\textwidth]{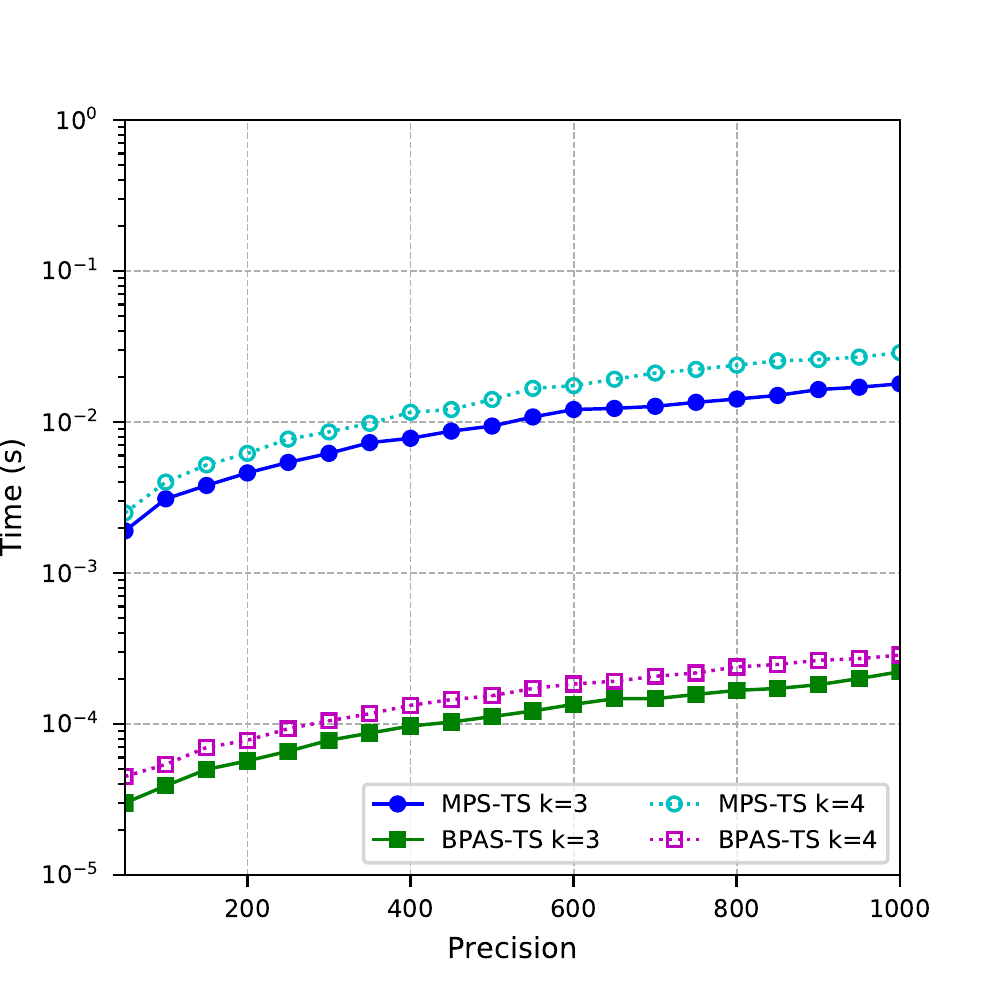}
		\caption{Computing ${\TS}(f, 1)$ \newline for $f = \prod_{i=1}^{k}({X_{2}}-i) + X_1({X_{2}}^{k-1}+{X_{2}})$.} 
		\label{fig:ts}
	\end{minipage}
\end{figure}

We also compare the factorization via Hensel's lemma and Taylor shift algorithms 
for a set of UPoPS $f = \prod_{i=1}^{k}(X_2-i) + X_1(X_2^{k-1}+X_2)$
in $\K[\![X_1]\!][X_2]$
with $k = 3, 4$ in Figures \ref{fig:fact} and \ref{fig:ts}.
Our factorization implementation is orders of magnitude faster than that of
{\tt RCPS}. However, factorization performs worse than expected 
compared to {\BPAS}, having already seen comparable performance of Weierstrass preparation 
in Figures~\ref{fig:wp-t1} and \ref{fig:wp-t2}.
This difference can be attributed to 
Taylor shift, the other core operation of {\HF}, as seen in Figure~\ref{fig:ts}.
The implementation in {\tt MPS} is slower than the same procedure 
in {\BPAS} by several order of magnitude. 
This, in turn, can be attributed to using {\Maple} matrix arithmetic, 
rather than the direct manipulation of C-arrays as in {\BPAS}, within
the Taylor shift algorithm.

\section{Conclusions and Future Work}
\label{sec:conc}
Throughout this work we have discussed the object-oriented design and implementation of 
power series and univariate polynomials over power series 
following lazy evaluation techniques. 
Basic arithmetic operations for both are examined as well as 
Weierstrass Preparation Theorem, Taylor shift, and factorization via Hensel's lemma 
for univariate polynomials over power series. 
Our implementation in {\Maple} is orders of magnitude faster than the existing 
multivariate power series implementation in the \texttt{PowerSeries} package of the {\RegularChains} library.
Moreover, our implementation is comparable 
with the C implementation of power series and univariate polynomials over power series in {\BPAS}.

Further work is needed to extend lazy evaluation techniques to more sophisticated algorithms.
For example, 
a general Extended Hensel Construction (EHC) \cite{Marc18LecNote}, and
the Abhyankar-Jung Theorem \cite{parusinski2012abhyankar}.
As a consequence, it is possible to re-implement the
EHC algorithm found in {\tt RegularChains} using this library.
Further, as {\Maple} supports multithreading,
it is possible to apply parallel processing to our algorithms. 
In particular, the computation of UPoPS coefficients in Weierstrass preparation 
is embarrassingly parallel.
Meanwhile, the successive application of Weierstrass preparation and Taylor shift 
in {\HF} present an opportunity for \textit{pipelining}.
Both should be exploited in to achieve even further 
performance improvements. 

\subsection*{Acknowledgements}
The authors would like to thank MITACS of Canada
(award IT19704)
and NSERC of Canada (award CGSD3-535362-2019).

	\addcontentsline{toc}{chapter}{Bibliography}
	\bibliographystyle{splncs04}
	\bibliography{references}{}
\end{document}